\documentclass[epsf,preprint,aps]{revtex4}

\usepackage{epsfig}

\def\be#1{\begin{equation}\label{#1}}
\def\ee{\end{equation}}
\def\bea#1{\begin{eqnarray}\label{#1}}
\def\eea{\end{eqnarray}}
\def\sp{\hspace{.5em}}
\def\Eq#1{Eq.(\ref{#1})}
\def\no{\nonumber \\}
\def\Lp{\Lambda p}
\def\Lpv{\mathbf{p}_\Lambda}
\def\pv{\mathbf{p}}
\def\adag{a^\dagger}
\def\acdag{a^{c\dagger}}
\def\mbf#1{\mbox{\boldmath $#1$}}
\def\Fig#1{Fig.(\ref{#1})}
\def\half{\frac{1}{2}}
\def\eps{\epsilon^\mu}
\def\epspm{\epsilon^\mu_\pm}
\def\epspp{\epsilon^\mu_+(\pv)}
\def\epsmp{\epsilon^\mu_-(\pv)}

\begin{document}

\title{Lorentz Invariance of Entanglement}
\author{Paul M. Alsing}
\email{alsing@ahpcc.unm.edu}
\affiliation{Albuquerque High Performance Computing Center,\\
University of New Mexico, Albuquerque, NM}
\author{Gerard J. Milburn}
\email{milburn@physics.uq.edu.au}
\affiliation{Centre for Quantum Computer Technology, \\
University of Queensland, Brisbane, Australia}
\date{\today}

\begin{abstract}
We study the transformation of maximally entangled states
under the action of Lorentz transformations in a fully relativistic setting.
By explicit calculation of the Wigner rotation,
we describe the relativistic analog of the Bell states
as viewed from two inertial frames moving
with constant velocity with respect to each other. Though the finite dimensional
matrices describing the Lorentz transformations are non-unitary,
each single particle state of the entangled pair undergoes an effective,
momentum dependent, local unitary rotation, thereby preserving the entanglement fidelity
of the bipartite state.
The details of how these unitary transformations are manifested are explicitly worked
out for the Bell states comprised of massive
spin $1/2$ particles and massless photon polarizations. The relevance of this work to
non-inertial frames is briefly discussed.
\end{abstract}

\maketitle


\section{Introduction}
Entanglement of bipartite quantum states forms a vital resource for many quantum
information processing protocols, including quantum teleportation, cryptography,
computation and clock synchronization. According to the principle of special relativity
the physics involved in utilizing such states should not depend on the arbitrary inertial
coordinate system from which the states are observed. Therefore we should expect the
states to transform unitarily from one inertial frame to another. This is clearly the case for
rotations. However, from the famous theorem by Wigner \cite{wigner} the finite dimensional representations
of Lorentz boosts are non-unitary. At first glance, it is then not immediately obvious where unitarity arises
in the case of boosts. The resolution to this apparent dilemma arises from the fact that
in relativistic quantum mechanics the creation and annihilation operators, as well
as the associated mode functions, for the quantum field that
creates a given state transform under Lorentz transformations (LTs) by local unitary spin-$j$
representations of the 3D rotation group \cite{weinberg95}. Key to these
transformations is the representation of the \textit{Wigner rotation} $W$,
which is a rotation in the rest frame of the particle, that leaves the
rest momentum invariant. The purpose of this work is to review
the role the Wigner rotation plays in restoring unitarity in the transformations between
relativistic single and multi-particle states. In particular, through explicit calculation of the
Wigner rotation we describe the observation of the entangled Bell states from two inertial
frames moving with constant velocity with respect to each other. The details are worked
out in a fully relativistic framework for the two important cases of spin entangled and
photon polarization entangled Bell states,
which occur most often in quantum information processing protocols. The end results of these calculations
will be that under Lorentz transformations, each constituent particle of
the relativistic generalization of the Bell states will undergo
an effective, momentum dependent, local unitary rotations, which will therefore preserve
the entanglement fidelity of the bipartite state.

For the purpose of concreteness we consider the symmetric Bell state
in the center of momentum frame $S$, with the two constituent particles
$A$ (Alice) and $B$ (Bob) travelling along the $\pm z$ direction.
For the symmetric Bell state
$\beta^{(1/2)}_{00}=(|\uparrow\,\uparrow\rangle+|\downarrow\,\downarrow\rangle)/\sqrt{2}$,
composed of spin $1/2$ electrons with the quantization axis along $z$, we will show that
an observer $S'$ travelling with constant velocity with respect to $S$ will observe
a rotation of the spins in the direction
of boost, at an angle less than the direction of the new spatial momentum.
For the photon polarization entangled state
$\beta^{(1)}_{00}(|HH\rangle+|VV\rangle)/\sqrt{2}$, where $H$ and $V$ represent horizontal and
vertical polarizations, we will find $S'$ observes a rotation
of the plane of polarization, tilted towards the direction
of boost, and perpendicular to the new observed momentum.
Though these two cases are analogous, the explicit details are
different due to the form of the \textit{little group} \cite{weinberg95} which
governs the invariance of the rest momentum for massive and
massless particles. For the massive electrons, the little group
governing the Wigner rotation is $SO(3)$, the ordinary group of
$3D$ rotations. For the massless photons, the little group is
$ISO(2)$, the Euclidean group of rotations and translations in the
$2D$ plane perpendicular to the propagation direction. Things are
a little more complicated in the case of massless particles,
since the little group in this case can induce gauge transformations
in the $4$-potentials. In order to ensure that the $4$-potentials
transform unitarily under boosts, we adopt the procedure of Han \textit{et al}
\cite{kim85}, which reduces to the choice of a particular gauge in which
the photon polarization vectors lie in the same plane as the electric
and magnetic fields. Though some generality is lost by gauge-fixing, the
explicit unitarity for the representations of the Lorentz boosts is sufficient gain
for most all quantum optical information processing applications.

The organization of this paper is as follows. In Section
\ref{qfields} we review the formalism of quantum
fields in Minkowski space, the representations of the Lorentz
transformations and Wigner's little group. In Section \ref{spinhalf}
specialize our discussion to the case of the electron Bell state $\beta^{(1/2)}_{00}$
and work out the Wigner rotation and transformed state for
a representative boost in a direction orthogonal to particle's momentum. In  Section \ref{spin1} we repeat
the previous calculations for the photon Bell state $\beta^{(1)}_{00}$.
Here we make special note of the work by Han \textit{et al.} \cite{kim85} which
shows how a LT on the polarization vectors, preceeded by a gauge transformation
leads to a pure rotation, which is finite dimensional and unitary. Both the
gauge transformation and the rotation are elements of the little group for photons.
Finally, in the last section we summarize our results and comment on their relevance
to the discussion of entanglement in non-inertial, accelerated frames.

\section{Quantum Fields in Minkowski Space}
\label{qfields} For our discussion of quantum fields in Minkowksi
space, we follow the text by Weinberg \cite{weinberg95}
and (for ease of reference) adopt his notation, metric signature and index ordering. As
such, Greek indices $\mu,\nu,$ etc. run over the four
spacetime coordinates labels $\{1,2,3,0\}$ with $x^0$ the time
component. Latin indices $i,j,k,$ etc.  run over the
three spatial coordinates labels  $\{1,2,3\}$. The spacetime
metric $\eta_{\mu\nu}$ is diagonal with elements $\{1,1,1,-1\}$.
Four-vectors are in un-boldfaced type while spatial vectors are
boldfaced. For e.g. the 4-momentum for particle of mass $m$ is
given by $p^\mu = (p^1,p^2,p^3,p^0) = (\mathbf{p},p^0)$, with norm
$p^\mu p_\mu = \mathbf{p}^2 - (p^0) = -m^2$. We use natural units
where $\hbar=c=1$, and occasionally include explicit factors of $c$ for clarity.

\subsection{Single Particle States}
\label{SingleParticle} Single particle quantum states are
classified by their transformation under the inhomogeneous Lorentz
group, or Poincar\'{e} group, consisting of homogeneous Lorentz
transformations (rotations and boots) $\Lambda$ and translations
$b$ (\cite{weinberg95}, Chapter 2). A general Poincar\'{e} transformation relates the coordinates
$x^\mu$  in an inertial frame $S$ to those of another inertial
frame $S'$ with coordinates $x^{'\mu}$ via
\be{1}
x^{'\mu} \equiv T(\Lambda,b)x^\mu = \Lambda^\mu_{\sp\nu} x^\nu + b^\mu.
\ee
For future reference, we denote the transformed 4-momentum
as $p' \to \Lp$ and its 3-vector spatial momentum as $\Lpv$.
A product of Lorentz transformations satisfies the composition rule
\be{2}
T(\bar{\Lambda},\bar{b}) \, T(\Lambda,b) = T(\bar{\Lambda}\Lambda,\bar{\Lambda}b+\bar{b}).
\ee

Single particle quantum states are denoted by $\Psi_{p,\sigma}$
where $p$ labels the 4-momenta and $\sigma$ labels all other
degrees of freedom. For our purposes, we may concentrate
on the spin degree of freedom; spin for massive particles and helicity
for massless particles. The state-vectors $\Psi_{p,\sigma}$ have
the property $P^\mu \Psi_{p,\sigma} = p^\mu \Psi_{p,\sigma}$, where
$P^\mu$ is the momentum operator and $p^\mu$ is its eigenvalue.
A Poincar\'{e} transformation $T(\Lambda,a)$ induces a linear unitary
transformation on the vectors in the physical Hilbert space of states via
\be{3}
\Psi\to U(\Lambda,b)\Psi.
\ee
The unitary operators $U(\Lambda,b)$ satisfy the same composition rule
as in \Eq{2} (with $T$ replaced by $U$). The commutation relations for the
Poincar\'{e} algebra \cite{weinberg95} tell us that under translations the
state-vectors transform as $U(1,b)\Psi_{p,\sigma} = e^{-ip\cdot b} \Psi_{p,\sigma}$.
Under homogeneous Lorentz transformations (LTs) $\Lambda$, the state-vector
$\Psi_{p,\sigma}$ with momentum $p$ must transform to a linear
combination of the state-vectors  $\Psi_{\Lp,\sigma}$ with momentum $\Lp$, i.e.
\be{4}
U(\Lambda)\Psi_{p,\sigma} = \sum_{\sigma'} C_{\sigma'\sigma}(\Lambda,p)\Psi_{\Lp,\sigma'}.
\ee
The matrix $C_{\sigma'\sigma}$ can be chosen to be block diagonal in the index $\sigma$,
with each block forming an irreducible representation of the inhomogeneous Lorentz group.

\subsubsection{Massive Particles}
\label{massive}
Consider for the moment the case of massive particles, $p^2<0$. We can always
choose some standard 4-momentum $k^\mu$ (usually taken in the particle's
rest frame) and express any $p^\mu$ of this class by
\be{5}
p^\mu = L^\mu_{\sp\nu}(p) \,k^\nu \qquad \textrm{or} \qquad  p=L(p)\,k,
\ee
where $L^\mu_{\sp\nu}(p)$ is some standard Lorentz transformation that depends on
$p$ and takes $k\to p$.  We can then define the state-vectors  $\Psi_{p,\sigma}$
in terms of standard momentum states $\Psi_{k,\sigma}$ as
\be{6}
\Psi_{p,\sigma} \equiv N(p) U(L(p)) \Psi_{k,\sigma'},
\ee
where $N(p)$ is a normalization factor which Weinberg conventionally takes as
$N(p) = \sqrt{k^0/p^0}$. Now the importance of the Wigner rotation can be
seen to arise as follows. Using the fact $U(L_1)U(L_2)=U(L_1 L_2)$
where $L_1$ and $L_2$ are arbitrary LTs, we have
upon acting on \Eq{6} with an arbitrary LT, $U(\Lambda)$
\bea{7}
U(\Lambda)\Psi_{p,\sigma} &=& N(p)\,U(\Lambda L(p))\,\Psi_{k,\sigma'} \no
&=&  N(p)\,U(L(\Lp))\, [\,U(L^{-1}(\Lp)\Lambda L(p))\,]\, \Psi_{k,\sigma'} \no
&\equiv& N(p)\,U(L(\Lp))\, U(W(\Lambda,p)) \, \Psi_{k,\sigma'}.
\eea
In the second line of \Eq{7} we have inserted the identity matrix
in the form of  $L(\Lp)\,L^{-1}(\Lp)$ in the argument of $U$ and
have defined the \textit{Wigner rotation} as the product of LTs in the argument of $U$ in the
square brackets:
\be{8}
 W(\Lambda,p) \equiv  L^{-1}(\Lp)\,\Lambda \,L(p).
\ee That $W$ is a rotation can be seen as follows from \Eq{8}.
Operating from right to left, $L(p)$ takes the standard momentum
$k$ to $L(p)k = p$. The LT, $\Lambda$ takes $p$ to $\Lp$. The final
LT, $L^{-1}(\Lp)$ takes $\Lp$ back to $k$. Thus
$W$ belongs to the subgroup of the homogeneous Lorentz group that
leaves $k^\mu$ invariant:
\be{9}
W^\mu_{\sp\nu} k^\nu = k^\mu.
\ee
This subgroup is called (Wigner's) \textit{little group}. The end
product of all this is that we can rewrite \Eq{4} as
\be{10}
U(\Lambda)\Psi_{p,\sigma} = \sqrt{\frac{(\Lp)^0}{p^0}}
\sum_{\sigma'} D_{\sigma'\sigma}(W(\Lambda,p))\Psi_{\Lp,\sigma'},
\ee
where $D(W)$ furnishes a representation of the little group element $W$.

For massive particles $p^2 = -m^2 <0, p^0>0$, the standard momentum can be
be chosen as $k^\mu = mc(0,0,0,1)$ and the little group is the usual group
of ordinary rotations in 3D, $SO(3)$. In this case the
$D^{(j)}_{\sigma',\sigma}(W(\Lambda,p))$ form the usual spin-$j$ representations
of the rotation group. The standard boost in the direction
$\hat{\mathbf{p}} \equiv \mathbf{p}/|\mathbf{p}|$ with rapidity $\eta$ defined by the relations
\be{11}
 \cosh\eta = \sqrt{\pv^2 + m^2}/m, \qquad \sinh\eta = |\pv|/m
\ee
is given by
\bea{12}
L^i_j(\eta) &=& \delta_{ij} + (\cosh\eta -1)\,\hat{p}_i \,\hat{p}_j, \no
L^i_0(\eta) &=& L^0_i(\eta) = \sinh\eta\, \hat{p}_i, \\
L^0_0(\eta) &=& \cosh\eta. \nonumber
\eea
The boost in \Eq{12} can always be written in the form
\be{13}
L(p) = R(\hat{\pv}) B_z(|\pv|) R^{-1}(\hat{\pv})
\ee
where $R(\hat{\pv})$ is a rotation that takes the $z$-axis into $\pv$ by first
rotating about the $y$-axis by an angle $\theta$ and then about the $z$-axis
by an angle of $\phi$, and $B_z(|\pv|)$ is a pure boost in the $z$-direction.
The unitary representation of $R(\hat{\pv})$ on the Hilbert space
is given by $U(R(\hat{\pv})) = e^{i\phi J_z} e^{i\theta J_y}$. Finally, if the LT, $\Lambda$
is a pure arbitrary 3D rotation $\cal{R}$, then $W(\Lambda,p)\equiv \cal{R}$ for all $p$.

\subsubsection{Massless Particles}
\label{massless} For massless particles $p^2=0, p^0>0$, the
standard momentum can be taken to be $k^\mu = (0,0,1,1)$. The
little group $W(\Lambda,p)$ which leaves this $k$ invariant (i.e. satisfies
\Eq{9}) is the group $ISO(2)$ which consists of
rotations $R_z(\theta)$ about the $z$-axis by an angle $\theta$ and
2D translations $S(\alpha,\beta)$ in the $x-y$ plane with
displacements vector $(\alpha,\beta,0,0)$. The Wigner rotation can be
expressed as a product of this rotation and translation as
\be{14}
W(\theta,\alpha,\beta)= S(\alpha,\beta) R_z(\theta).
\ee
This leads to a representation of $D(W)$ as \cite{weinberg95}
\be{14a}
D_{\sigma'\sigma}(W) = e^{i\theta\sigma}\,\delta_{\sigma'\sigma},
\ee
where
$\sigma$ labels the possible helicity states of the particle. In
the case of photons, $\sigma=\pm 1$ corresponding to states of right
and left circularly polarization. Instead of \Eq{10}, we now have
the transformation of state-vectors under a homogeneous Lorentz
transformation given by
\be{15}
U(\Lambda)\Psi_{p,\sigma} =
\sqrt{\frac{(\Lp)^0}{p^0}} e^{i\sigma\theta(\Lambda,p)} \,
\Psi_{\Lp,\sigma},
\ee
where $\theta(\Lambda,p)$ is defined by
\Eq{14}.  Due to the gauge freedom in the electromagnetic field,
there is still work that needs to performed to construct the
Wigner rotation for photons and compute the angle $\theta$ in
\Eq{15}. We take this up in Section \ref{spin1}.

\subsection{Multi-Particle States}
\label{multi-particle} The generalization of the single particle
states to many particle states is relatively straight forward but
notationally cumbersome. We denote a multi-particle state-vector by
$\Phi_{p_1,\sigma_1,n_1;p_2,\sigma_2,n_2;\ldots}$ where $p_i$ labels the
momentum, $\sigma_i$ is the spin $z$-component
(or helicity for massless particles), and $n_i$ is a species label for the
$i$th particle. In keeping with the notation of \cite{weinberg95}, $\Phi$ could
refer to either free particle states $\Psi$, or 'In' and 'Out' scattering states.
We will be concerned only with free particle states, but will retain
Weinberg's notation of using $\Phi$ for the state-vectors and $U(\Lambda)\to U_0(\Lambda)$
to denote the representations of Lorentz transformations on the Hilbert space of states. From now
own we will be concerned mainly with proper orthochronous LTs.

A multi-particle state transforms as the direct product of single
particles states. Considering massive particles for the time being, we can write
the transformation of a multi-particle state under a proper orthocrhonous
inhomogeneous Lorentz transformation $U_0(\Lambda,b)$ as
\bea{16} \lefteqn{U_0(\Lambda,b)
\Phi_{p_1,\sigma_1,n_1;p_2,\sigma_2,n_2;\ldots} =
\exp\Big((-ib_\mu (p_1^\mu + p_2^\mu + \cdots)\Big) } \no
& & \times \sqrt{\frac{(\Lp_1)^0(\Lp_2)^0\cdots}{p_1^0p_2^0\cdots}}
  \sum_{\sigma'_1 \sigma'_2 \cdots}
  D^{(j_1)}_{\sigma'_1\sigma_1}\Big(W(\Lambda,p_1)\Big)
  D^{(j_2)}_{\sigma'_2\sigma_2}\Big(W(\Lambda,p_2)\Big)\cdots  \no
& & \times \Phi_{\Lp_1,\sigma'_1,n_1;\Lp_2,\sigma'_2,n_2;\ldots} \quad .
\eea

The $0$-particle state $\Phi_0$ is the Lorentz invariant vacuum with normalization
of unity, $(\Phi_0, \Phi_0) = 1$, where the parentheses denote the inner product
on the Hilbert space.  The $1$-particle state is denoted by $\Phi_q$, where we
use the shorthand notation $q=(\pv,\sigma,n)$ to represent the relevant quantum numbers.
This has norm $(\Phi_{q'},\Phi_q)=\delta(q'-q)$$\equiv \delta(\pv'-\pv) \delta_{\sigma'\sigma} \delta_{n'n}$.
The $2$-particle state $\Phi_{q'q}$ is physically equivalent to the state $\Phi_{qq'}$
so we must take its norm to be
$(\Phi_{q'_1 q'_2} \Phi_{q_1 q_2})$ $= \delta(q'_1-q_1)\delta(q'_2-q_2)\pm \delta(q'_2-q_1)\delta(q'_1-q_2)$,
where the $-$ is taken if both particles are fermions and $+$ otherwise.
The general $N$-particle state $\Phi_{q_1 q_2 \dots q_N}$ is taken to have norm
$(\Phi_{q'_1 q'_2 \dots q'_M},\Phi_{q_1 q_2 \dots q_N}) =$
$\delta_{NM}\sum_{\cal{P}}\delta_{\cal{P}}$ $\prod_i\delta(q_i-q'_{{\cal{P}}_i})$ where the sum is
over all signed permutations of the integers $\{1,2,\ldots,N\}$.

The above multi-particle states can be produced by the action of the \textit{creation operator}
$\adag(q)$ which adds adds a particle with quantum numbers $q$ to the front of the list of
of particles in the state, $\adag(q)\Phi_{q_1 q_2\ldots q_n}= \Phi_{q q_1 q_2\ldots q_n}$.
The general $N$-particle state can be produced from the vacuum by acting upon it with
$N$ creation operators
\be{17}
\adag(q_1) \adag(q_2) \ldots \adag(q_N) \Phi_0 = \Phi_{q_1 q_2\ldots q_N}.
\ee
Our main point of interest is the observation that in order for the state in
\Eq{17} to transform properly, i.e. in accordance with \Eq{16}, the creation
operator must satisfy the transformation rule
\bea{18}
U_0(\Lambda,b) \adag(\pv\sigma n) U^{-1}_0(\Lambda,b) &=&
e^{-i(\Lp)\cdot b} \sqrt{(\Lp)^0/p^0}  \no
&  \times & \sum_{\sigma'} D^{(j_n)}_{\sigma'\sigma}(W(\Lambda,p))\,\adag(\Lpv\sigma' n),
\eea
where $j_n$ is the spin of the $n$th particle species. (For massless particles, the
$D$ \Eq{18} must be replaced by that in \Eq{14a}).

We can now create quantum fields $\psi_l(x) = \psi^+_l(x) + \psi^-_l(x)$ where the
$\pm$ indicates the positive and negative frequency field operators and $l$ is
the field index label, e.g. $l=\{1,2,3,4\}$ for a spin-$1/2$ Dirac bispinor representing the
electron-positron field, and $l\to\mu=\{1,2,3,0\}$ for spin-$1$ electromagnetic
$4$-potential field. The positive frequency annihilation field $\psi^+_l(x)$ and
negative frequency creation field $\psi^-_l(x)$ are given by
\bea{19}
\psi^+_l(x) &=& \sum_{\sigma n} \int \,d^3p \, u_l(x;\pv,\sigma,n) \, a(\pv,\sigma,n), \\
\psi^-_l(x) &=& \sum_{\sigma n} \int \,d^3p \, v_l(x;\pv,\sigma,n) \, \adag(\pv,\sigma,n),
\eea
where the mode functions $u_l(x;\pv,\sigma,n)$ and $v_l(x;\pv,\sigma,n)$ are chosen so
that under LTs each field is multiplied by a position-independent matrix
\bea{20}
U_0(\Lambda,b) \psi^+_l(x) U^{-1}_0(\Lambda,b) &=&
             \sum_{l'} D_{l l'}(\Lambda^{-1}) \psi^+_l(\Lambda x + b), \\
\label{20a}
U_0(\Lambda,b) \psi^-_l(x) U^{-1}_0(\Lambda,b) &=&
             \sum_{l'} D_{l l'}(\Lambda^{-1}) \psi^-_l(\Lambda x + b).
\eea Here the $D_{l l'}(\Lambda)$ are matrices which form a block
diagonal representation of the LTs for the fields, with each block
containing irreducible representations.

If we now form $U_0(\Lambda,b) \psi^+_l(x) U^{-1}_0(\Lambda,b)$
from \Eq{19} and use the adjoint of \Eq{18}, we obtain consistency
with \Eq{20} and \Eq{20a} if the following transformation of the mode functions
holds for massive particles (see \cite{weinberg95} for the details)
\be{21}
\sum_{\sigma'} u(\Lpv,\sigma',n)
D^{(j_n)}_{\sigma'\sigma}\Big(W(\Lambda,p)\Big) =
\sqrt{\frac{p^0}{(\Lp)^0}}\sum_l D_{l'l}(\Lambda)
u_l(\pv,\sigma,n),
\ee
and
\be{22}
\sum_{\sigma'}
v(\Lpv,\sigma',n) D^{(j_n)*}_{\sigma'\sigma}\Big(W(\Lambda,p)\Big)
= \sqrt{\frac{p^0}{(\Lp)^0}}\sum_l D_{l'l}(\Lambda)
v_l(\pv,\sigma,n).
\ee
and for massless particles (from \Eq{15})
\be{21a}
u(\Lpv,\sigma,n)\, e^{i\sigma\theta(\Lambda,p)} =
\sqrt{\frac{p^0}{(\Lp)^0}}\sum_l D_{l'l}(\Lambda)
u_l(\pv,\sigma,n)
\ee
and
\be{22a}
v(\Lpv,\sigma,n)\, e^{-i\sigma\theta(\Lambda,p)} =
\sqrt{\frac{p^0}{(\Lp)^0}}\sum_l D_{l'l}(\Lambda)
v_l(\pv,\sigma,n)
\ee
In obtaining \Eq{21}-\Eq{22a} we have
used the fact that under pure translations, $U_0(1,b)$ one can deduce
that the mode functions must take the form
$u_l(x;\pv,\sigma,n)=(2\pi)^{-3/2}e^{ip\cdot x}u_l(\pv,\sigma,n)$
and $v_l(x;\pv,\sigma,n)=(2\pi)^{-3/2}e^{-ip\cdot
x}v_l(\pv,\sigma,n)$.

We can interpret \Eq{21} as follows. Recall
that $u_{l'}(\pv,\sigma,n)$ forms a column vector of field
components characterized by a momentum $\pv$, spin or helicity
$\sigma$ and species index $n$ which we temporarily denote as
$\vec{u}(\pv,\sigma,n)$. For electrons, $\vec{u}$ has four
bispinor components and $\sigma=\pm 1/2$ denote spin up or down
along some quantization axis. For photons, $\vec{u}$ has four
spacetime components ($l'\to\mu$) and $\sigma=\pm 1$ denote states
of right and left circularly polarization.  Under a Lorentz
transformation $U(\Lambda,b)$, $\vec{u}(\pv,\sigma,n)$ is
transformed to a new vector $\vec{u}'(\Lpv,\sigma,n)$.
\Eq{21} tells us that we can compute the
transformed vector $\vec{u}'(\Lpv,\sigma,n)$ in two ways. Up to a
normalization factor, the right hand side of \Eq{21} indicates that (in
matrix notation) we can compute $\vec{u}'(\Lpv,\sigma,n) =
D(\Lambda) \vec{u}(\pv,\sigma,n)$ for a fixed spin or helicity
index $\sigma$, i.e. by transforming the field components
according to $D(\Lambda)$. The left hand side of \Eq{21} states,
that for a fixed field component $l'$, we can
re-write  $\vec{u}'(\Lpv,\sigma,n)$ as a linear combination of the
spin/helicity mode functions with momentum $\Lpv$ with
coefficients given by the spin-$j_n$ matrix representations
$D^{(j_n)}_{\sigma'\sigma}(W)$ of the Wigner rotation $W$.
It is in this later case that we see that a Lorentz transformation
induces a momentum dependent, local unitary rotation of the spin components of each
particle in a multi-particle state. Each constituent single particle state
is transformed at most into a superposition of spin states with the
transformed momentum $\Lpv$. Such local unitary rotations cannot
effect the entanglement fidelity of the multi-particle state.
We shall give explicit examples in the next section.

Though we will mainly be concerned in this paper with the transformation
of states, for completeness we list the corresponding transformations of
the creation and annihilation operators in the new
inertial frame ($x' = \Lambda x + b$). These can re-expressed from
\Eq{18} (using the unitarity of the rotation matrices
$D^{(j_n)}_{\sigma'\sigma}(W)$) as
\bea{23}
\lefteqn{U_0(\Lambda,b) \, a(\pv,\sigma,n) \, U^{-1}_0(\Lambda,b) =
e^{i(\Lp)\cdot b} \sqrt{(\Lp)^0/p^0} }\no
& & \times
\sum_{\sigma'} D^{(j_n)}_{\sigma\sigma'}(W^{-1}(\Lambda,p)) \,
a(\Lpv,\sigma',n),
\eea
and
\bea{24} \lefteqn{U_0(\Lambda,b) \,
\adag(\pv,\sigma,n)  \, U^{-1}_0(\Lambda,b) = e^{-i(\Lp)\cdot b}
\sqrt{(\Lp)^0/p^0} }\no
& & \times \sum_{\sigma'}
D^{(j_n)*}_{\sigma\sigma'}(W^{-1}(\Lambda,p)) \, \adag(\Lpv,\sigma',n),
\eea

In the next two sections we will specialize the results of the Lorentz transformation
rules for the mode functions \Eq{21} and \Eq{21a}, to the specific cases
of $2$-qubit spin entangled states and $2$-qubit polarization entangled states.

\section{Electrons: Spin 1/2 Fields}
\label{spinhalf}
The spin $1/2$ Dirac field is given by
\be{25}
\psi_l(x) = \sum_{\sigma} \int \,d^3p \, \left[ u_l(\pv,\sigma) \, e^{i p\cdot x} \, a(\pv,\sigma)
+ v_l(\pv,\sigma) \, e^{-i p\cdot x} \, \acdag(\pv,\sigma)\right],
\ee
where we have dropped the species label $n$. Here $a(\pv,\sigma)$ annihilates a particle in
the (Dirac bi-)spinor state $u_l(\pv,\sigma)$, corresponding to an electron with momentum $\pv$ with
spin $\sigma=\pm 1/2$ along a quantization axis, which we shall take as the $z$-axis.
The charge conjugate creation operator (needed
to conserve electric charge) $\acdag(\pv,\sigma)$
creates antiparticles in the spinor state $v_l(\pv,\sigma)$.
Since $e^{i p\cdot x} = e^{i(-E t + \pv \cdot \mathbf{x})}$, the factor $e^{-i p\cdot x}$ associated with
the antiparticle state $v_l(\pv,\sigma)$ implies that it can also be interpreted
as a negative energy solution with negative momentum.

The mode functions of momentum $\pv$
and spin $\sigma=\pm 1/2$ are given by Lorentz transformations
\be{26}
u_l(\pv,\sigma) = \sqrt{\frac{mc}{p^0}}\,D\Big(L(p)\Big) u(0,\sigma),
\qquad v_l(\pv,\sigma) = \sqrt{\frac{mc}{p^0}}\,D\Big(L(p)\Big) v(0,\sigma)
\ee
of their rest frame values which are taken to be \cite{weinberg95}
\be{27}
u(0,1/2) = \frac{1}{\sqrt{2}}
\left[
\begin{array}{c}
  1 \\
  0 \\
  1 \\
  0
\end{array}
\right], \qquad
u(0,-1/2) = \frac{1}{\sqrt{2}}
\left[
\begin{array}{c}
  0 \\
  1 \\
  0 \\
  1
\end{array}
\right],
\ee
\be{28}
v(0,1/2) = \frac{1}{\sqrt{2}}
\left[
\begin{array}{c}
  0 \\
  1 \\
  0 \\
 -1
\end{array}
\right], \qquad
v(0,-1/2) = \frac{1}{\sqrt{2}}
\left[
\begin{array}{c}
 -1 \\
  0 \\
  1 \\
  0
\end{array}
\right].
\ee
The mode functions $u(\pv,\sigma)$ and $v(\pv,\sigma)$ are eigenvectors of
$-i p^\mu \gamma_\mu$ with eigenvalues $+1$ and $-1$ respectively, i.e.
\be{29}
(i p^\mu \gamma_\mu + m)u(\pv,\sigma)=0, \qquad (-i p^\mu \gamma_\mu + m)v(\pv,\sigma)=0,
\ee
so that the field \Eq{25}, satisfies the Dirac equation
\be{30}
( \gamma^\mu \partial_\mu + m) \psi(x) = 0.
\ee
In the above we have used the notation of Wienberg \cite{weinberg95} to define the
gamma matrices as follows:
\be{31}
\{\gamma^\mu, \gamma^\nu\} = 2 \eta^{\mu\nu}, \quad \gamma^0 \equiv -i \alpha_4 =
-i\left(
\begin{array}{cc}
  \mbf{0}\sp & \mbf{1} \\
  \mbf{1}\sp & \mbf{0}
\end{array}
\right), \quad
\mbf{\gamma}\equiv -i \mbf{\alpha} = -i
\left(
\begin{array}{cc}
  \mbf{0} & \mbf{\sigma} \\
  -\mbf{\sigma} & \mbf{0}
\end{array}
\right),
\ee
where $\mbf{\sigma} = (\sigma_1,\sigma_2,\sigma_3)$ are the usual $2\times 2$ Pauli matrices.
The above is the \textit{chiral} representation in which $\gamma^5\equiv -i\gamma^0\gamma^1\gamma^2\gamma^3$
(which commutes with each of the $\gamma^\mu$) is diagonal,
$$
\gamma^5 =\left(
\begin{array}{cc}
  \mbf{1} & \mbf{0} \\
  \mbf{0} & -\mbf{1}
\end{array}
\right).
$$

For an infinitesimal Lorentz transformation with parameters $\omega^\mu_{\sp\nu}$,
written in the form
\be{32}
\Lambda^\mu_{\sp\nu} = \delta^\mu_{\sp\nu} + \omega^\mu_{\sp\nu},
\ee
the induced unitary transformation $D(\Lambda)$ of the spinors is given by
\be{33}
D(\Lambda) = 1 + \frac{i}{2} \omega_{\mu\nu} \mathcal{J}^{\mu\nu}.
\ee
Here the generators of the Lorentz transformations on the spinors is given by
$\mathcal{J}^{\mu\nu} = -i/4\, [\gamma^\mu, \gamma^\nu]$. In the chiral representation,
these matrices take the explicit form
\be{34}
\mathcal{J}^{ij} = \frac{1}{2} \epsilon_{ijk}
\left(
\begin{array}{cc}
  \sigma_k & 0 \\
  0 & \sigma_k
\end{array}
\right), \qquad
\mathcal{J}^{i0} = \frac{i}{2}
\left(
\begin{array}{cc}
  \sigma_i & 0 \\
  0 & -\sigma_i
\end{array}
\right),
\ee
where the matrices on the left generate rotations and the matrices on the right generate boosts.
Note that the generators $\mathcal{J}^{ij}$ are Hermitian so that rotations are represented by
unitary matrices. However, the generators are $\mathcal{J}^{i0}$ are anti-Hermitian and therefor
pure boosts are not represented by unitary matrices. This follows from the well know theorem
that all finite dimensional representations of boost matrices are non-unitary \cite{wigner}.

The relativistic two-particle state $\Phi(\beta^{(1/2)}_{00})$ associated with the
non-relativistic spin entangled Bell state
$\beta^{(1/2)}_{00}=(|\uparrow\,\uparrow\rangle+|\downarrow\,\downarrow\rangle)/\sqrt{2}$
is given by
\bea{35}
\Phi(\beta^{(1/2)}_{00}) &\equiv& \frac{1}{\sqrt{2}}
\left(
\Phi_{\pv,1/2;-\pv,1/2} + \Phi_{\pv,-1/2;-\pv,-1/2}
\right) \\
&=&\frac{1}{\sqrt{2}}
\Big(
u_A(\pv,1/2) \otimes u_B(-\pv,1/2) + u_A(\pv,-1/2) \otimes u_B(-\pv,-1/2)
\Big).
\eea
The state $\Phi(\beta^{(1/2)}_{00})$ represents two particles $A$ (Alice) and $B$ (Bob)
travelling in opposite directions (which we take to be the $z$-direction)
with equal and opposite momenta $\pv$ in a superposition of products states of both
spins up and both spins down, along a quantization axis which, without loss
of generality, we also take as the $z$-axis.
There are two things to note here. First, if we had made a unitary transformation $\mathcal{U}$
given by
\be{36}
\mathcal{U} =
\frac{1}{\sqrt{2}}\left(
\begin{array}{cc}
  \mbf{1} & \mbf{1} \\
  \mbf{1} & -\mbf{1}
\end{array}
\right),
\ee
and defined new rest frame spinors $\tilde{u}(0,\sigma) = \mathcal{U} \,u(0,\sigma)$ and
$\tilde{v}(0,\sigma) = \mathcal{U} \, v(0,\sigma)$ we'd find $\tilde{u}(0,1/2) = (1,0,0,0)$ and
$\tilde{u}(0,-1/2) = (0,1,0,0)$ while $\tilde{v}(0,1/2) = (0,0,0,1)$ and $\tilde{v}(0,-1/2) = (0,0,-1,0)$.
The upper two "large" components of $\tilde{u}(0,\sigma)$ correspond to familiar
2-spinor components which we associate with the non-relativistic states $|\uparrow\rangle$ and
$|\downarrow\rangle$ while the lower two "small" components of $\tilde{u}(0,\sigma)$
are of order $|\pv| c/E$, and are typically neglected in the non-relativistic theory.
This is true even when we consider the boosted states $\tilde{u}(\pv,\sigma)$.
In the chiral representation used here, the states $u$ and $v$ are just rotated versions
of $\tilde{u}$ and $\tilde{v}$. Thus in the
limit of small velocities, the state $\Phi(\beta^{(1/2)}_{00})\to \beta^{(1/2)}_{00}$.

Secondly, the two-particle state $\Phi(\beta^{(1/2)}_{00})$ involves only the single-particle states $u(\pv,\sigma)$,
and not the anti-particle states $v(\pv,\sigma)$. This occurs because the positive and negative
energy states $u$ and $v$ transform among themselves separately and do not mix with each other
under proper LTs, as well as under spatial inversions \cite{BD}. The
factor $e^{i p\cdot x}$ associated with $u$ is future-directed in the light cone in $p$ space
and the factor $e^{-i p\cdot x}$ associated with $v$ is past-directed. Since $p\cdot x$ is
a Lorentz invariant, the positive and negative energy states remain distinct, and hence do
not mix.

Our two-particle state $\Phi(\beta^{(1/2)}_{00})$ transforms as superposition of direct
product states according to \Eq{16}, so it is enough for us to consider the Lorentz transformation
of the of the single particle state $u(\pv,\sigma)$. Our goal is to find the Wigner rotation
$W(\Lambda,p)$ \Eq{8}, associated with an arbitrary Lorentz boost $\Lambda$ of the state $u(\pv,\sigma)$.
$W$ is a rotation that keeps the standard momentum $k^\mu = mc(0,0,0,1)$ invariant, \Eq{9}.
Without loss of generality we take $p^\mu = L(p)^\mu_{\sp\nu} \, k^\mu$, with $L(p)$ a standard boost
given by \Eq{12}, along the $z$-axis with rapidity $\eta$:
\be{37}
L(p) = \left(
\begin{array}{cccc}
  1\sp & 0\sp & 0 & 0 \\
  0\sp & 1\sp & 0 & 0 \\
  0\sp & 0\sp & \cosh\eta & \sinh\eta \\
  0\sp & 0\sp & \sinh\eta & \cosh\eta
\end{array}
\right), \qquad
p^\mu = L(p)^\mu_{\sp\nu} \, k^\mu = mc
\left[
\begin{array}{c}
  0 \\
  0 \\
  \sinh\eta \\
  \cosh\eta
\end{array}
\right]
\ee
Recall that the rows and column of $L(p)$ and $p^\mu$ are labelled by indices $(1,2,3,0)$.
In \Eq{37} we have made a boost to a coordinate system $S'$ travelling in the $-z$ direction with
velocity given by $\tanh(-\eta) = v/c$ so that in $S'$ the particle, initially at rest in $S$
with momentum $k^\mu$ and state $u(0,\sigma)$,
will be observed to have velocity $v/c$ in the $+z$ direction  with state $u(\pv,\sigma)$ \Big(where
$|\pv| = \gamma_v m v$ with $\gamma_v \equiv (1-v^2/c^2)^{-1/2}$\Big).

For a Lorentz transformation in the $\pm z$ direction, $W$ in \Eq{8} trivially reduces to the
identity matrix ($0$ angle rotation), since two boosts in the same direction are equivalent to
a single boost along the same direction. Thus, as observed from either Alice's or Bob's rest frame,
the state remains unaltered. Therefore, without loss of generality, we will consider
a boost $\Lambda$ in the $x$ direction with rapidity $\omega$ corresponding to a LT to a frame travelling along
the $-x$ direction with velocity $-v_x$ such that $\tanh(-\omega) = v_x/c$:
\be{38}
\Lambda = \left(
\begin{array}{cccc}
  \cosh\omega & 0\sp  & 0\sp  & \sinh\omega \\
  0           & 1\sp  & 0\sp  & 0 \\
  0           & 0\sp  & 1\sp  & 0 \\
  \sinh\omega & 0\sp  & 0\sp  & \cosh\omega
\end{array}
\right),
\ee
with
\be{39}
(\Lambda p)^\mu = \Lambda^\mu_{\sp\nu} \, p^\mu = mc
\left[
\begin{array}{c}
  \sinh\omega \cosh\eta \\
  0 \\
  \sinh\eta \\
  \cosh\omega \cosh\eta
\end{array}
\right] \equiv mc
\left[
\begin{array}{c}
 \sin\theta \sinh\xi \\
  0 \\
 \cos\theta \sinh\xi \\
  \cosh\xi
\end{array}
\right].
\ee
In \Eq{39} we have introduced the polar angle $\theta$ which $\Lpv$ makes with respect to the $z$ axis in
the $xz$ plane, and the rapidity $\xi$ by the relations
\bea{40}
\tan\theta &=& \frac{\sinh\omega}{\tanh\eta} = (\Lpv)_1/(\Lpv)_3, \\
\cosh\xi   &=& \cosh\omega \cosh\eta = E_{\Lpv}/mc^2, \\
\sinh\xi   &=& \sqrt{\cosh^2\omega \cosh^2\eta-1} = |\Lpv|/mc.
\eea

We now want to construct the standard boost Lorentz transformation $L^{-1}(\Lambda p)$ such that
$L(\Lambda p)$ takes $k\to\Lambda p$ directly from rest. From \Eq{12} we identify
$(\hat{p}_\Lambda)_1 = \sin\theta$ and $(\hat{p}_\Lambda)_3 = \cos\theta$ and the rapidity as $\xi$
appropriate for $L(\Lambda p)$. For $L^{-1}(\Lambda p)$ we let $\theta\to\theta + \pi$ (a LT in the reverse direction)
thereby obtaining
\be{41}
L^{-1}(\Lambda p) = \left(
\begin{array}{cccc}
  1 + (\cosh\xi -1) \sin^2\theta\sp       & 0\sp   & (\cosh\xi -1) \sin\theta \cos\theta\sp   & -\sin\theta \sinh\xi \\
  0\sp                                    & 1\sp   & 0\sp                                     & 0 \\
  (\cosh\xi -1) \sin\theta \cos\theta\sp   & 0\sp   & 1 + (\cosh\xi -1) \sin^2\theta \sp       & -\cos\theta \sinh\xi  \\
 -\sin\theta \sinh\xi\sp                  & 0\sp   & -\cos\theta \sinh\xi \sp                 & \cosh\xi
\end{array}
\right),
\ee
A brute force calculation reveals that indeed, $\L^{-1}(\Lambda p)^\mu_{\sp\nu}\, (\Lambda p)^\nu = k^\mu$.
A quick way to see this is to note that the $4$th column of $\L(\Lambda p)$ \Big(obtained from
\Eq{41} by letting $(\sin\theta,0,\cos\theta) \to (-\sin\theta,0,-\cos\theta)$\Big) is just $(\Lambda p)/mc$
given by \Eq{39}. Since $\L^{-1}(\Lambda p) \, \L(\Lambda p) = I$ by construction, $\L^{-1}(\Lambda p)$
acting on the $4$th column of $\L(\Lambda p)$ produces $k^\mu = mc(0,0,0,1)$.

\begin{figure}[htb]
\centering
\epsfig{file=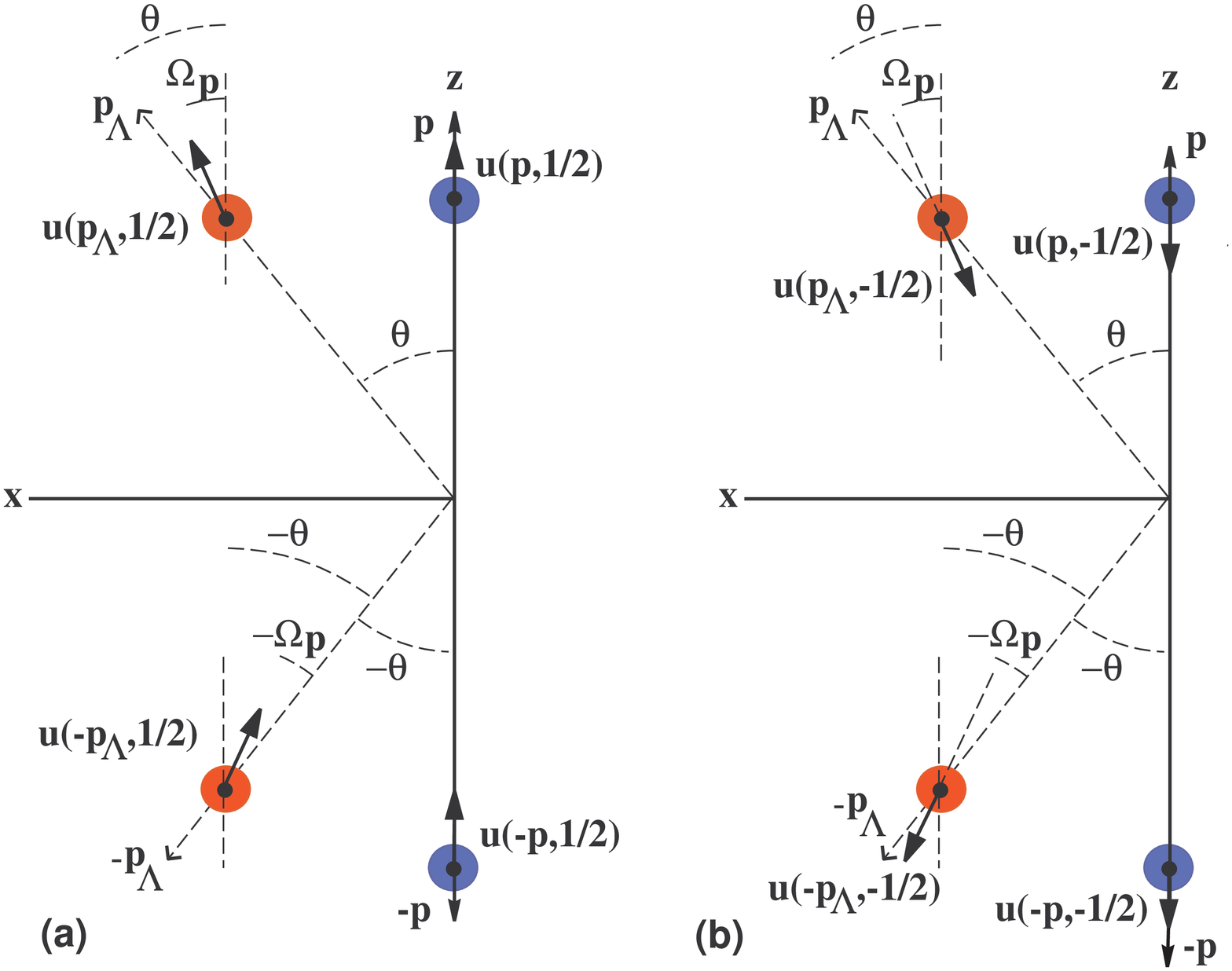,height=4.5in,width=6in}
\caption{Effect of a boost $\Lambda$ in x direction on the electron spinors $u(\pm\pv,\sigma)$.
In the frame $S$, the (blue) electrons (a) $u(\pm\pv,1/2)$, (b) $u(\pm\pv,-1/2)$
are travelling in the $\pm z$ direction with momentum $\pm\pv$
with spins aligned or anti-aligned along the quantization axis $z$.
The figures show the (red) electrons (a) $u(\pm\Lpv,1/2)$, (b) $u(\pm\Lpv,-1/2)$  as observed in a frame $S'$
travelling along the $-x$ direction with respect to $S$ with velocity $v_x/c$.
As observed by $S'$, the momentum $\pm\Lpv$ of the electrons rotates by an angle $\pm\theta$
about the $+y$ axis (pointing out of the plane of the page)
where $+\theta$ is a counter clockwise rotation. However,
the direction of spin is observed by $S'$ to rotate
by an angle $\pm\Omega_{\pv}$, in the same sense as $\theta$, but of lesser magnitude.
}\label{electrons}
\end{figure}

In order to calculate the Wigner rotation $W(\Lambda,p) = \L^{-1}(\Lambda p)\,\Lambda\,L(p)$,
we need the product of the matrices $\Lambda\,L(p)$ :
\be{42}
\Lambda\,L(p) =
\left(
\begin{array}{cccc}
  \cosh\omega \sp & 0 \sp & \sinh\omega\sinh\eta \sp   & \sinh\omega \cosh\eta \\
  0\sp            & 1 \sp & 0 \sp                      & 0 \\
  0\sp            & 0 \sp & \cosh\eta \sp              & \sinh\eta \\
  \sinh\omega\sp  & 0 \sp & \cosh\omega \sinh\eta \sp  & \cosh\omega \cosh\eta
\end{array}
\right).
\ee
To check that $W$ represents a pure rotation we consider a spatial vector $z^\mu \equiv (0,0,1,0)$
in the rest frame and compute its transformation under $W$. For a pure rotation we must have
\be{43}
(Wz)^\mu = W^\mu_{\sp\nu}\,z^\nu \equiv
\left[
\begin{array}{c}
  \sin\Omega_{\pv} \\
  0 \\
  \cos\Omega_{\pv} \\
  0
\end{array}
\right], \qquad \textrm{where} \sp\sp z^\mu \equiv
\left[
\begin{array}{c}
  0 \\
  0 \\
  1 \\
  0
\end{array}
\right].
\ee
Equation (\ref{43}) represents a pure rotation about the $y$ axis by an angle $\Omega_{\pv}$,
since the two pure boosts in \Eq{42} both occur in the $xz$ plane.
Noting that $\Lambda \,L(p) z$ is the third column of \Eq{42},
a straight forward, but tedious calculation of $\L^{-1}(\Lambda p)$ times this vector,
using the definitions of $\xi$ and $\theta$ from \Eq{40} yields
\be{44}
(Wz)^\mu =
\left[
\begin{array}{c}
  \sinh\eta \sinh\omega/ (1+\cosh\omega \cosh\eta) \\
  0 \\
  (\cosh\omega + \cosh\eta) /(1+\cosh\omega \cosh\eta) \\
  0
\end{array}
\right],
\ee
allowing us to identify the Wigner rotation angle $\Omega_{\pv}$ by
\be{45}
\tan\Omega_{\pv} = \frac{\sinh\eta \sinh\omega}{\cosh\omega + \cosh\eta} \equiv
\frac{\sinh\eta \tanh\eta}{\cosh\omega + \cosh\eta} \, \tan\theta.
\ee
From \Eq{45} we can infer that for all values of $\eta$ and $\omega$ associated with
boosts $L(p)$ in the $z$ direction and with $\Lambda$ in the $x$ direction, respectively we have
\be{46}
\Omega_{\pv} < \theta, \qquad  0 \leq \eta,\; \omega < \infty,
\ee
where $\theta$ is the angle that $\Lpv$ makes with $\pv$ (see \Fig{electrons}).
%
%

We are now ready to describe the effect of this Wigner rotation on the transformations
of the spinor $u(\pv,\sigma)$ according to \Eq{21}.
First we need a representation of $u(\pv,\sigma)$ for arbitrary $\pv$. This is given
by the formula \Eq{26} using the rest frame spinors in \Eq{27}. From \Eq{32} - \Eq{34}
we have the spinor representation $D\Big(L(p)\Big)$ of a standard boost $L(p)$ of rapidity
$\zeta$
\bea{47}
\lefteqn{D\Big(L(p)\Big) = e^{i/2\mathcal{J}^{0i}\omega_{0i}} =
\exp\left[-\frac{\zeta}{2}
\left(
\begin{array}{cc}
  \mbf{\sigma}\cdot\mbf{\hat{p}} & \mbf{0} \no
  \mbf{0}& -\mbf{\sigma}\cdot\mbf{\hat{p}}
\end{array}
\right) \right]}\\
& = &
\cosh\zeta/2 \left(
\begin{array}{cccc}
  1-\hat{p}_3 \tanh\zeta/2 & -\hat{p}_- \tanh\zeta/2 & 0 & 0 \\
  -\hat{p}_+ \tanh\zeta/2 & 1+\hat{p}_3 \tanh\zeta/2 & 0 & 0 \\
  0 & 0 & 1+\hat{p}_3 \tanh\zeta/2 & \hat{p}_- \tanh\zeta/2 \\
  0 & 0 & \hat{p}_+ \tanh\zeta/2 & 1-\hat{p}_3 \tanh\zeta/2
\end{array}
\right).
\eea
In \Eq{47}, $L(p)$ is a coordinate Lorentz transformation to a frame $S'$ moving
with velocity $v/c = |\pv| c/ E_{\pv} = \tanh(-\zeta)$ such that from $S'$ the particle at
rest in frame $S$ is observed to have velocity $v/c$. The vector
$\mbf{\hat{p}} = (\hat{p}_1, \hat{p}_2,\hat{p}_3)$ is a unit vector in the direction of $\pv$
with $\hat{p}_\pm \equiv \hat{p}_1 \pm i \hat{p}_2$ and $\omega_{0i}=\hat{p}_i$.
In terms of the transformed momenta $\pv$ and energy $E_{\pv}$ we have the following relations
\bea{48}
\cosh\zeta = \frac{E_{\pv}}{mc^2}, & \qquad & -\sinh\zeta = \frac{|\pv|}{mc}, \qquad -\tanh\zeta = \frac{v}{c} \no
\cosh\zeta/2 = \sqrt{\frac{E_{\pv}+mc^2}{2mc^2}} & \qquad & -\tanh\zeta/2 = \frac{|\pv|c}{E_{\pv}+mc^2}.
\eea
Taking into account that $p^0 = mc\cosh\zeta$ so that $\sqrt{mc/p^0}=\sqrt{mc^2/E_{\pv}}$,
\Eq{26} yields
\be{49}
u(\pv,1/2) = \frac{\cosh\zeta/2}{\sqrt{2\cosh\zeta}}
\left[
\begin{array}{c}
  1-\hat{p}_3 \tanh\zeta/2 \\
  -\hat{p}_+ \tanh\zeta/2 \\
  1+\hat{p}_3 \tanh\zeta/2 \\
  \hat{p}_+ \tanh\zeta/2
\end{array}
\right], \quad
u(\pv,-1/2) = \frac{\cosh\zeta/2}{\sqrt{2\cosh\zeta}}
\left[
\begin{array}{c}
  -\hat{p}_- \tanh\zeta/2  \\
   1+\hat{p}_3 \tanh\zeta/2\\
  \hat{p}_- \tanh\zeta/2 \\
   1-\hat{p}_3 \tanh\zeta/2
\end{array}
\right].
\ee

The content of \Eq{21} is that under a Lorentz transformation $\Lambda$ taking $p\to\Lambda p$
the transformed spinors \Big(right hand side of \Eq{21}\Big) can be re-written as a Wigner rotation
of the spinors $u(\Lpv,\sigma)$ \Big(left hand side of \Eq{21}\Big), the later of which can
be obtained from \Eq{49} by a substitution of $\mbf{\hat{p}} \to \mbf{\hat{p}_{\Lambda}}$
with the appropriate redefinition $\cosh\zeta \to E_{\Lpv}/mc^2$. With the Wigner angle
$\Omega_{\pv}$ in hand, the rotation matrices on the left hand side of \Eq{21} are given
by \cite{edmonds}
\be{50}
D^{(j_n)}_{\sigma'\sigma}\Big(W(\Lambda,p)\Big) =
\left(
\begin{array}{cc}
  \cos(\Omega_{\pv}/2) & -\sin(\Omega_{\pv}/2) \\
  \sin(\Omega_{\pv}/2) & \cos(\Omega_{\pv}/2)
\end{array}
\right),
\ee
with the rows and columns of the matrix in \Eq{50} labelled by $\sigma = (1/2,-1/2)$.
Thus in matrix notation we can write \Eq{21} as
\bea{51}
u'(\pv,\half) &\equiv& \sqrt{\frac{p^0}{\Lambda p^0}} \, D(\Lambda) u(\pv,\half) =
\cos\left(\frac{\Omega_{\pv}}{2}\right) \, u(\Lpv,\half) + \sin\left(\frac{\Omega_{\pv}}{2}\right) \, u(\Lpv,-\half), \\
u'(\pv,-\half) &\equiv& \sqrt{\frac{p^0}{\Lambda p^0}} \, D(\Lambda) u(\pv,-\half) =
-\sin\left(\frac{\Omega_{\pv}}{2}\right) \, u(\Lpv,\half) + \cos\left(\frac{\Omega_{\pv}}{2}\right) \,u(\Lpv,-\half).
\eea
Note that for $u(-\pv,\sigma)$ the standard boost in \Eq{37} is performed in the opposite direction (coordinate
transformation to a frame moving along the $+z$ axis with velocity $v/c$) so that we simply change
the sign of the rapidity $\eta\to -\eta$. This leads to the following sign changes
\be{52}
\pv \to -\pv \quad \Rightarrow \quad \theta \to -\theta, \qquad \Omega_{-\pv} = - \Omega_{\pv}.
\ee
In \Fig{electrons} we illustrate the transformation of the product states $u_A(\pv,1/2) \otimes u_B(-\pv,1/2)$
and $u_A(\pv,-1/2) \otimes u_B(-\pv,-1/2)$ appearing as terms in $\Phi(\beta^{1/2}_{00})$,
which correspond to the non-relativistic product states $|\uparrow_A,\uparrow_B\rangle$
and $|\downarrow_A,\downarrow_B\rangle$, respectively. The effect of the Lorentz boost $\Lambda$ is
to rotate $\pv \to \Lpv$ through and angle $\theta$ defined by \Eq{40}. The orientations of the spins
with respect to the quantization axis $z$ are rotated by the momentum dependent Wigner angle $\Omega_{\pv}$ defined in
\Eq{45}, such that $\Omega_{\pv}<\theta$. The rotation is counter-clockwise for particles momentum $\pv$
and clockwise for particles with momentum $-\pv$.

\section{Photons: Spin 1 Fields}
\label{spin1}
The massless spin $1$ photon field is given by
\be{53}
a^\mu(x) =  \int\frac{d^3p}{\sqrt{(2\pi)^3\,2 p^0}}
\,\sum_{\sigma=\pm 1}\left[ \eps(\pv,\sigma) \, e^{i p\cdot x} \, a(\pv,\sigma)
+ \eps(\pv,\sigma)^* \, e^{-i p\cdot x} \, \adag(\pv,\sigma)\right],
\ee
where $\adag(\pv,\sigma)$ creates photons in $\sigma=\pm 1$ helicity states (right and left
circular polarization) $\eps(\pv,\sigma)$. Since \Eq{53} has the form of a $4$-vector field
the gauge independent representation $D(\Lambda)$ of Lorentz transformation $\Lambda$ is given by the
LT itself \cite{weinberg95}, i.e.
\be{54}
x^{'\mu} = \Lambda^\mu_{\sp\nu} \,x^\nu \quad \Rightarrow \quad
\epsilon^{'\mu}(\pv,\sigma) \equiv  D(\Lambda)^\mu_{\sp\nu} \, \epsilon^\nu(\pv,\sigma)
= \Lambda^\mu_{\sp\nu} \, \epsilon^\nu(\pv,\sigma).
\ee
However, as is well known, $a^\mu(x)$ cannot be a pure $4$-vector field since the electromagnetic
field has only two degrees of freedom. Thus we have a $4$-vector field
with a gauge freedom. Matters are also complicated by the fact that while rotations of
$4$-vectors are represented by finite dimension unitary matrices, the finite dimension
matrices representing boosts are non-unitary. The question at hand is can one find a
finite dimensional \textit{unitary} representation for the transformation of the polarization vectors?
This was answered by Han \textit{et al} \cite{kim85} who showed that by first pre-multiplying a polarization
vector by a matrix $\mathcal{D}$ in the little group appropriate for photons, and then applying the boost,
the net effect is a pure spatial rotation of $\epsilon^{\mu}(\pv,\sigma)$.
This procedure essentially reduces to a choice of a particular gauge (described below) in which there
are only two photon polarization vectors \cite{note} which always lie in the plane perpendicular to the
photon's momentum.  This choice of gauge consistent with most common definitions of
polarization vectors found in the quantum optics literature.
In the following we follow gauge-fixing choice of \cite{kim85}, and afterwards return
to make connection with the gauge independent transformation
equation for massless particles as given by Weinberg, \Eq{15}.

\begin{figure}[htb]
\centering
\epsfig{file=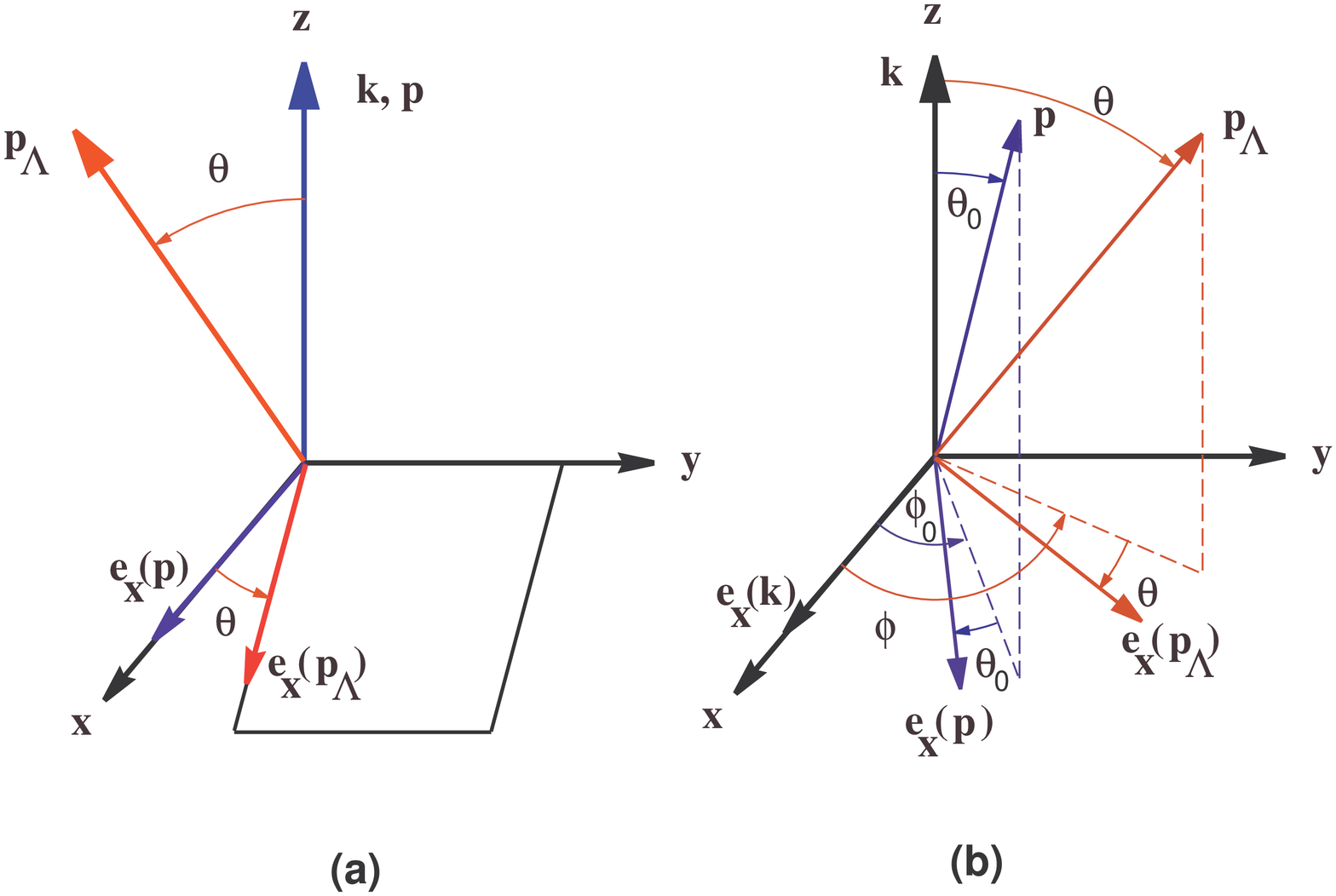,height=3.5in,width=6.0in}
\caption{(a) Effect of pure-boost $\Lambda=B_x(\omega)$ in x direction on the polarization
vector $\epsilon_x(\pv)$ given by the $3$-vector portion of
$\big(\epsilon^\mu_+(\pv)+ \epsilon^\mu_+(\pv)\big)/\sqrt{2}$ .
In a frame $S$ the photon (blue) is
propagating in the $+z$ with momentum $\pv$ with the orthogonal polarization vector
$\epsilon_x(\pv)$. In the frame $S'$, travelling in the $-x$
direction with respect to $S$ with velocity $v_x/c$, the photon
(red) is observed to have momentum $\Lpv$ inclined at a polar angle
$\theta$ in the $+xz$ plane. The plane of polarization of $S$ is
observed by $S'$ to rotate to a new plane at angle $\theta$ with
respect to $S$. (b) Effect of an arbitrary Lorentz transformation $\Lambda$.
In frame $S$ (blue) the photon has momentum $\pv$, not necessarily along the $z$ direction,
with orthogonal polarization vector $\epsilon_x(\pv)$. In a frame $S'$, related to $S$ by a Lorentz transformation
$\Lambda$, the triad $(\epsilon_x(\pv), \epsilon_y(\pv),\pv)$ in $S$ is observed to be
rigidly rotated to the triad $(\epsilon_x(\Lpv), \epsilon_y(\Lpv),\Lpv)$.}\label{photons}
\end{figure}

Our ultimate goal is to describe the effect of Lorentz boost on the $2$-qubit polarization
entangled state
$\beta^{(1)}_{00}=(|H\,H\rangle+|V\,V\rangle)/\sqrt{2}$
which is given by
\bea{55}
\Phi(\beta^{1}_{00}) &\equiv& \frac{1}{\sqrt{2}}
\left(
\Phi_{\pv,1;-\pv,1} + \Phi_{\pv,-1;-\pv,-1}
\right) \\
&=&\frac{1}{\sqrt{2}}
\Big(
\eps_A(\pv,1) \otimes \eps_B(-\pv,1) + \eps_A(\pv,-1) \otimes \eps_B(-\pv,-1)
\Big).
\eea
Here we have again taken Alice and Bob to be travelling along the $z$ axis with equal
and opposite momentum $\pv$. We let the single-particle horizontal polarization state $|H\rangle$
be represented by  the positive helicity state $\eps(\pv,+1) \equiv \epspp$, and the vertical polariztion
state $|V\rangle$ by $\eps(\pv,-1)\equiv \epsmp$. Again, it is enough to consider the transformation
of the single particle polarization state $\eps(\pv,\sigma)$. For the first calculation we
choose the photon momentum $\pv$ in $S$ to lie along the $z$ axis, the direction of the standard
momentum, and consider a boost along the $x$ direction.
The net result of this calculation will be the simple result, that in the boosted frame, an observer $S'$
travelling along the $-x$ axis will observe a tilting of the plane of polarization towards the $+x$ axis,
(see \Fig{photons}a).
We then generalize this calculation to an arbitrary LT (not necessarily a boost only) for
a photon of momentum $\pv$ along an arbitrary direction in $S$. We show that for the observer $S'$,
the triad of $3$-vectors $(\mathbf{e_x}(\pv), \mathbf{e_y}(\pv),\pv)$ is rigidly rotated to the triad
$(\mathbf{e_x}(\Lpv), \mathbf{e_y}(\Lpv),\Lpv)$, where $\mathbf{e_x}(\pv)$ is the $3$-vector portion of
$\Big(\epsilon^\mu_+(\pv)+ \epsilon^\mu_+(\pv)\Big)/\sqrt{2}$ and $\mathbf{e_y}(\pv)$ is the $3$-vector portion of
$-i\,\Big(\epsilon^\mu_+(\pv)- \epsilon^\mu_+(\pv)\Big)/\sqrt{2}$
(see \Fig{photons}b).

We begin by considering a photon travelling in the $+z$ direction in the local
inertial frame $S$. We take as the standard momentum
$k^\mu = (0,0,1,1)$ such that $k^\mu k_\mu = 0$. An arbitrary $4$-potential has the form
$A^\mu(x) = A^\mu \exp\Big(i k(z-ct)\Big)$ with $A^\mu = (A^1, A^2, A^3, A^0) = (\mbf{A},A^0)$. As stated in
Section \ref{qfields}, the little group for photons is $ISO(2)$ \Big(often called $E(2)$\Big),
the Euclidean group of rotations and translations in the polarization plane perpendicular to
the momentum of the photon. The generator of rotations $J_3$ is given by \cite{kim85}
\be{56}
J_3 =
\left(
\begin{array}{cccc}
  0\sp & -i\sp & 0\sp & 0\sp \\
  i\sp & 0\sp  & 0\sp & 0\sp \\
  0\sp & 0\sp  & 0\sp & 0\sp \\
  0\sp & 0\sp  & 0\sp & 0\sp
\end{array}
\right),
\ee
called the helicity operator. The generators $A$ and $B$ for translations in the plane of polarization
are given by $A=J_2 + K_1$ and $B=-J_1 + K_2$ with
$[J_3,A]=i\,B$, $[J_3,B]=-i\,A$ and $[A,B]=0$, where $J_i$ is a rotation about the
$i$th axis and $K_i$ is a pure boost along the $i$th direction \cite{weinberg95,kim82}.
The operators $A$ and $B$ generate translations and
their particular form need not concern us here. However, we note
that they are responsible for inducing gauge transformations of the $4$-potentials,
$A^{'\mu} = A^\mu + \partial^\mu \chi$. We take as our $4$-potentials $A^\mu$ eigenstates
of $J_3$ namely
\be{57}
J_3 \, \epspm(\mathbf{k}) = \pm \epspm(\mathbf{k}), \qquad \epspm(\mathbf{k}) = \frac{1}{\sqrt{2}}
\left[
\begin{array}{c}
  1 \\
  \pm i \\
  0 \\
  0
\end{array}
\right].
\ee

In order for a $A^\mu$ to be a proper $4$-potential (i.e. represent a physical polarization $4$-vector)
it must satisfy the following
two properties: (1) $A^0=0$ and (2) $\mbf{p}\cdot\mbf{A}=0$. These
two conditions are equivalent to the combined effect of the Lorentz condition
\be{58}
\frac{\partial}{\partial x^\mu} A^\mu(x) = p^\mu A^\mu(x) = 0,
\ee
and the transversality condition
\be{59}
\mbf{\nabla}\cdot\mbf{A}(\mbf{x})=0, \qquad \textrm{or} \qquad \mbf{p}\cdot\mbf{A}=0.
\ee
The first of these conditions \Eq{58} is a Lorentz invariant statement, the second
\Eq{59}, is not. Han \textit{et al} refers to these two conditions as the \textit{helicity gauge}.

From the form of $k^\mu$ and $\epspm(\mbf{k})$ in \Eq{57}, a standard boost $L(p)$ in the $z$ direction
with rapidity $\eta$, given by \Eq{37}, will change the momentum to
$p^\mu = L(p)^\mu_{\sp\nu}\,k^\nu = (0,0,k,k)$ with
$k\equiv |\pv|=(\cosh\eta + \sinh\eta)$ but will leave $\epspm(\pv)\equiv\epspm(\mathbf{k})$ invariant. This last
statement is obvious, since vectors perpendicular to the direction of a pure boost
are unaltered. Thus, as in the case for massive spin $1/2$ particles, the state observed from either
Alice's or Bob's frame of reference, is unaltered.
Therefore, we can again consider, without loss of generality, a boost
$\Lambda$ in the $x$ direction given by \Eq{38} i.e. a transformation to a frame $S'$ moving
in the $-x$ direction with velocity $v_x/c$ such $\tanh(-\omega) = v_x/c$. In the frame $S'$,
the photon, originally tavelling in the $+z$ direction in frame $S$ will be observed
to be travelling in the $+xz$ plane. Under $\Lambda$, $p \to \Lambda p$ with
\be{60}
(\Lambda p)^\mu = \Lambda^\mu_{\sp\nu}\,p^\nu =
k \, \left[
\begin{array}{c}
  \sinh\omega\\
  0 \\
  1 \\
  \cosh\omega
\end{array}
\right] =
k \cosh\omega \, \left[
\begin{array}{c}
  \tanh\omega\\
  0 \\
  1/\cosh\omega \\
  1
\end{array}
\right] \equiv
k \, \cosh\omega \left[
\begin{array}{c}
  \sin\theta\\
  0 \\
  \cos\theta \\
  1
\end{array}
\right].
\ee
In \Eq{60} we have defined the polar rotation angle $\theta$ that $\Lpv$ makes with $\pv$
by factoring out $|\Lpv|= k\cosh\omega$ and defining
\be{61}
\sin\theta \equiv \tanh\omega, \quad \cos\theta \equiv 1/\cosh\omega, \quad \tan\theta = \sinh\omega.
\ee
Consider
$\tilde{\epsilon}^\mu_\pm \equiv \Lambda\,\epspm(\pv) = k\,(\cosh\omega,\pm i, 0, \sinh\omega)$.
Although it satisfies the transversality condition $(\Lambda p)^\mu \tilde{\epsilon}_{\pm\mu}(\pv) = 0$,
it fails to be a valid $4$-potential since $\tilde{\epsilon}^0_\pm(\pv) \neq 0$.

In order to calculate the Wigner rotation $W(\Lambda,p) = L^{-1}(\Lambda p)\,\Lambda\,L(p)$ we note
the standard boost $L(p)$ which takes $k\to p$ can in general be written as
\be{62}
L(p) = R(\mathbf{\hat{p}}) \, B_z(|\pv|),
\ee
where $R(\hat{\mathbf{p}})$ is a pure rotation that takes the $z$ axis into $\mathbf{\hat{p}}$.
For an a momentum in an arbitrary direction
$\hat{\mathbf{p}} = (\sin\theta\cos\phi, \sin\theta\sin\phi, \cos\theta)$ we can take
$R(\hat{\mathbf{p}}) = R_z(\phi)\,R_y(\theta)$, where $R_y(\theta)$ is a rotation about
the $y$ axis taking $(0,0,1)$ to $(\sin\theta,0,\cos\theta)$, followed by a rotation
about the $z$ axis by the angle $\phi$, taking the intermediate direction to $\mathbf{\hat{p}}$.
$B_z(|\pv|)$ is a boost in the $z$ direction taking the standard momentum $k$ of unit magnitude
$|\mbf{k}|=1$ to magnitude $|\pv|$, given by
\be{63}
B_z(u) = \left(
\begin{array}{cccc}
  1\sp\sp & 0\sp\sp & 0 & 0 \\
  0\sp\sp & 1\sp\sp & 0 & 0 \\
  0\sp\sp & 0\sp\sp & (u^2+1)/2u & (u^2-1)/2u \\
  0\sp\sp & 0\sp\sp & (u^2-1)/2u & (u^2+1)/2u
\end{array}
\right).
\ee
In terms of a rapidity $\xi$ (see \Eq{37} with $\eta\to\xi$), we have $(u^2+1)/2u = \cosh\xi$ or $u = \cosh\xi + \sinh\xi$.
In addition, we define the polarization vector $\epspm(\pv)$ for arbitrary momentum $\pv$
in terms of standard polarization vector $\epspm(\mbf{k})$
of \Eq{57} by
\bea{64}
\epspm(\pv) &\equiv& L(p)\,\epspm(\mbf{k}) \no
&=& R(\mathbf{\hat{p}}) \, B_z(|\pv|)\,\epspm(\mbf{k}) = R(\mathbf{\hat{p}}) \,\epspm(\mbf{k}),
\eea
where the last equality follows since $\epspm(\mbf{k})$ is left invariant by boosts along the $z$ direction.

For the particular case we have chosen to consider, i.e. $\pv = |\pv|\,\mbf{\hat{k}}$ along the $z$ direction, we have
$L(p)\equiv B_z(k)$, with $k=|\pv|$.
With our Lorentz transformation taken to be $\Lambda \equiv B_x(\omega)$ \Eq{38}, we can compute
$L(\Lambda p)$ by substituting $\Lpv$ for $\pv$ in \Eq{62}, with $|\Lpv|=k\cosh\omega$.
Using the angle $\theta$  defined in \Eq{60} which $\Lambda p = L(\Lambda p)\, k$ makes with the $z$ axis, we have
$L(\Lambda p) = R_y(\theta) \, B_z(|\Lpv|)$ with
\be{65}
R_y(\theta) =
\left(
\begin{array}{cccc}
  \cos\theta \sp & 0\sp & \sin\theta \sp& 0\sp \\
  0\sp & 1\sp & 0\sp & 0\sp \\
  -\sin\theta \sp& 0 \sp& \cos\theta \sp& 0 \sp\\
  0\sp & 0\sp & 0\sp & 1\sp
\end{array}
\right).
\ee
Note that we can write $B_z(|\Lpv|) = B_z(|\Lpv|/|\pv|)\,B_z(|\pv|) = B_z(\cosh\omega)\,L(p)$ so that
we can write $(\Lambda p)$ in two equivalent forms:
\be{66}
(\Lambda p) = B_x(\omega) \; p = R_y(\theta) \, B_z(\cosh\omega) \; p = L(\Lambda p)\; k.
\ee

Collecting these results we have
\bea{67}
W(\Lambda,p) &=& L^{-1}(\Lambda p)\,\Lambda\,L(p) \no
&=& L^{-1}(p)\, \Big[ B_z^{-1}(\cosh\omega)\, R^{-1}_z(\theta) \, B_x(\omega) \Big] \, L(p) \no
& \equiv & L^{-1}(p)\, \mathcal{D}^{-1}(\omega) \, L(p),
\eea
where we have defined
\be{67.5}
\mathcal{D}(\omega) \equiv  B_x^{-1}(\omega) \, R_z(\theta) \, B_z(\cosh\omega).
\ee
A trivial rearrangement of second equality of \Eq{66} shows that
\be{68a}
\mathcal{D}(\omega)^\mu_{\sp\nu} \, p^\nu = p^\mu,
\ee
so that $\mathcal{D}(\omega)$ is a member of the little group of $p$, i.e. LTs which leave $p$
(as opposed to $k$) invariant. \Eq{68a} also arises from a rearrangement of the defining property
of the Wigner rotation \Eq{9}, using the expression for $W$ in \Eq{67} and $p = L(p)\,k$.
We also note $\mathcal{D}(\omega)$ induces gauge transformations when acting on $4$-potentials:
\bea{69}
\bar{\epsilon}^\mu_\pm(\pv) \equiv \mathcal{D}(\omega)^\mu_{\sp\nu} \epsilon^\nu_\pm(\pv) &= &
(1,\pm i, -\tanh\omega, -\tanh\omega) \no
&=& \epspm(\pv) - k^\mu \tanh\omega
\eea
(as can be shown by direct matrix multiplication)
which can be associated with a gauge function $\chi = (ct -z) \, k^\mu \, \tanh\omega $.

By construction, we can also transform $p\to (\Lambda p)$ via the product of matrices
$\Lambda\,\mathcal{D}(\omega)$ acting on $p$. Similarly, if we precede the action of $\Lambda$
on $\epspm(\pv)$ by $\mathcal{D}(\omega)$ we find \cite{kim85}
\bea{68}
\epsilon^{'\mu}_\pm(\pv) &\equiv& \Lambda \,\mathcal{D}(\omega) \epspm(\pv) \no
&=& \Big(\Lambda \, \Lambda^{-1} \Big) \, R_y(\theta) \, \Big(B_z(\omega)\,\epspm(\pv) \Big) \no
&=& R_y(\theta) \, \epspm(\pv)\no
&\equiv& \epspm(\Lpv)
\eea
where we have used the fact that $B_z(\cosh\omega)\,\epspm(\pv) = \epspm(\pv)$.
Since $\theta= \tan^{-1}(\sinh\omega)$ is the polar angle $(\Lambda p)$
makes with respect to $p$, the net effect of the transformation is just a rotation of the plane of
polarization by the angle $\theta$ \big(see \Fig{photons}a\big). Thus we have
$\epsilon^{'\mu}_\pm(\pv,\sigma) \equiv \epspm(\Lpv)$, i.e. the polarization vector
appropriate for a photon with momentum in the direction $\Lpv$ (see \Eq{64}\,).
Note that $\epspm(\Lpv)$ is a valid $4$-potential in the helicity gauge since $\epsilon^0_\pm(\Lpv)=0$ and
$(\Lambda p)^\mu \, \epsilon_{\pm\mu}((\pv) = 0$ as required by \Eq{58} and \Eq{59}.

The salient point here is that the representation $D(\Lambda)$ of the Lorentz transformation
$\Lambda$ as given by $\Lambda\,\mathcal{D}(\omega)$ induces the unitary rotation $R_y(\theta)$
on the polarization vector $\epspm(\pv)$ by the Wigner angle $\theta$. This derives from \Eq{66}
which states that $\Lambda p$ can be reached in two ways from $p$:
first by the direct action of $\Lambda=B_x$ on $p$, and second by a boost $B_z$ along the $z$
direction with rapidity $\cosh\omega$ acting on $p$,
followed by a Wigner rotation about the $y$ axis by the angle $\theta$.
These parameters are related by $|\Lpv|/|\pv|=\cosh\omega$
and $\tan\theta = \sinh\omega$.

We can now generalize the above arguments to an arbitrary Lorentz transformation $\Lambda$, which is
not necessarily a pure boost, and for momentum $\pv$ in $S$ which lies along an arbitrary direction.
Thus we take $p = L(p)\, k$ with $L(p)$ given by \Eq{62} and $\Lambda$ arbitrary.
The key ingredient is to find $\mathcal{D}$ which
leaves $p$ invariant. We begin by generalizing \Eq{66}
\bea{69}
(\Lambda p) &=& L(\Lambda p)\,k \no
\Lambda \, p &=& R(\mbf{\hat{p}_\Lambda}) \, B_z(|\mbf{\hat{p}_\Lambda}|) \, L^{-1}(p) \, p \no
\Rightarrow p &=& \Lambda^{-1}\, R(\mbf{\hat{p}_\Lambda}) \, B_z(|\mbf{\hat{p}_\Lambda}|) \, L^{-1}(p) \, p \no
&\equiv& \mathcal{D} \,p,
\eea
where we have used $L(\Lambda p) = R(\mbf{\hat{p}_\Lambda}) \, B_z(|\mbf{\hat{p}_\Lambda}|)$
and $k = L^{-1}(p) \, p$.
By using this definition of $L(\Lambda p)$ and  pre-multiplying by unity in the form of $L^{-1}(p)\,L(p)$ we obtain
\bea{70}
W(\Lambda,p) &=& L^{-1}(\Lambda p)\,\Lambda\,L(p) \no
             &=& L^{-1}(p)\, \mathcal{D}^{-1} \, L(p)
\eea
Then defining the representation $D(\Lambda)$ of the LT $\Lambda$ as
\be{71}
D(\Lambda) = \Lambda \, \mathcal{D},
\ee
as opposed to just $D(\Lambda) = \Lambda$, we have its action upon $\epspm(\pv)$
given by
\bea{72}
\epsilon^{'\mu}_\pm(\pv) &\equiv& \Lambda \,\mathcal{D} \epspm(\pv) \no
&=& \Big(\Lambda \, \Lambda^{-1} \Big) \, R(\mbf{\hat{p}_\Lambda}) \, \Big(B_z(\mbf{\hat{p}_\Lambda}|)\,
                                L^{-1}(p) \,\epspm(\pv)\Big) \no
&=& R(\mbf{\hat{p}_\Lambda})\, \epspm(\mbf{k}) \no
& \equiv & \epspm(\Lpv),
\eea
where the last equality follows from \Eq{64}.
The last two lines of \Eq{72} leads to the transformation
\be{73}
\epspm(\Lpv) = D(\Lambda) \, \epspm(\pv) = R(\mbf{\hat{p}_\Lambda})\,R^{-1}(\mbf{\hat{p}})\,\epspm(\pv)
\ee
which is explicitly unitary.
As depicted in \Fig{photons}b, the transformation in \Eq{73} rigidly rotates the triad of $3$-vectors
$(\mathbf{e_x}(\pv), \mathbf{e_y}(\pv),\pv)$ into the triad
$(\mathbf{e_x}(\Lpv), \mathbf{e_y}(\Lpv),\Lpv)$.

To make connection with Weinberg's transformation equation \Eq{21a}, we note that in
\Eq{68} and \Eq{73} there are no explicit phase factors of $\exp(i\,\sigma\theta)$.
This results from the (helicity) gauge-fixing convention of Han \textit{et al} which represents
the LT $D(\Lambda)$ acting on the $4$-potentials as $\Lambda\,\mathcal{D}$. In essence, the
gauge transformations induced by $\Lambda$ are undone by the $p$-little group element $\mathcal{D}$.
The price one pays for ensuring explicit unitary representations for boosts is that $D(\Lambda)$ must
be represented by $\Lambda \mathcal{D}$.

This is in contrast to Weinberg's gauge invariant representation of $D(\Lambda)$ by $\Lambda$ itself, \Eq{54}.
In the local Lorentz frame of the photon, the Wigner rotation is represented in a gauge invariant manner
by the product of a translation $S(\alpha,\beta)$ and a rotation $R_z(\theta)$, \Eq{14}. Acting upon
$\epspm(\mbf{k})$, the rotation about the $z$ axis produces the phase factor $\exp(i\,\sigma\theta)$
appearing in \Eq{21a}. However, the action of the translation $S(\alpha,\beta)$ induces gauge
transformations on the $4$-potential, so that in general the transformed potential contains
a non-zero time component $\epsilon^0_\pm$ and is no longer a helicity state. For general
momentum $\pv$ the transformed $4$-potential is given by $\Lambda^\mu_{\sp\nu} \, \epsilon^\nu_\pm(\pv)$
plus gauge induced components parallel to $p^\mu$ (see discussion in \cite{weinberg95}, p249-251).
All this stems from the requirement that $\epsilon^0_\pm(\pv)$ and hence the quantum field operator
$a^0$ vanish in all Lorentz frames, ensuring that the field $a^\mu$ cannot be a true $4$-vector field.
Of course, the gauge invariant physical electric and magnetic fields are not affected by such considerations.
However, it is the polarization vectors that are found in quantum optics to be most useful in representing
the state of the system. At the minor cost of losing some generality by gauge fixing, one gains explicit
unitarity in the representations of boosts by finite dimensional matrices.

\section{Summary and Discussion}
In non-relativistic quantum mechanics, the only kinematic
transformations of reference frames we are allowed to consider are translations and rotations, which are explicitly unitary.
In relativistic quantum mechanics, we must also consider Lorentz boosts, which when
represented by finite dimensional matrices are explicitly non-unitary.
In spite of this, each single particle state in a multi-particle state undergoes an effective,
momentum dependent, local unitary rotation under Lorentz boosts governed by the
little group element $W$ which leaves the appropriate standard momentum $k$ invariant.
For massive spin $1/2$ particles, the standard momentum (in the particle's rest frame)
is $k^\mu=mc(0,0,0,1)$ and the little group
is $SO(3)$, the group of ordinary rotations in 3D. Even though $W$ itself is not unitary,
its $3\times 3$ $(x,y,z)$-block acts as an effective rotation matrix (since the
components $W^t_i$ need not be zero).
For a pure boost
taking momentum $\pv$ into $\Lpv$, the spin of the transformed particle is rotated
by the Wigner angle $\Omega_{\pv}$, which is in the same sense, but less in magnitude
than the polar angle $\theta$ which $\Lpv$ makes with $\pv$.
For massless photons, the little group is $ISO(2)$, the group of rotations
and translations in the plane perpendicular to the standard momentum $k^\mu = (0,0,1,1)$.
Though $W$ itself is not unitary, in a gauge invariant description of the states,
its $2\times 2$ $(x,y)$-block acts as effective rotation matrix (since components outside
this block are not necessarily zero).
By fixing the choice of gauge
the transformation which takes the polarization vector $\epspm(\pv)$ to $\epspm(\Lpv)$
can be made explicitly unitary, i.e. a $4\times 4$ rotation matrix. In this case
the triad of $3$-vectors $(\epsilon_x(\pv), \epsilon_y(\pv),\pv)$ in one inertial frame is observed to be
rigidly rotated to the triad $(\epsilon_x(\Lpv), \epsilon_y(\Lpv),\Lpv)$ in
another inertial frame. Since a Lorentz transformation of a (massive or massless) multi-particle state
acts as a direct product, each constituent single particle state is transformed at most into
a superposition of spin or helicity states with the appropriate transformed momenta. Consequently,
tracing out over one state in maximally entangled bipartite state will still produce
a maximally mixed density matrix for the reduced state. The entanglement fidelity is not effected
by the Lorentz transformation. In this work we explicitly demonstrated the above considerations
for the relativistic generalization of a symmetric Bell state comprised of electrons, and of
photons, for arbitrary strength Lorentz boosts.
\begin{figure}[htb]
\centering
\epsfig{file=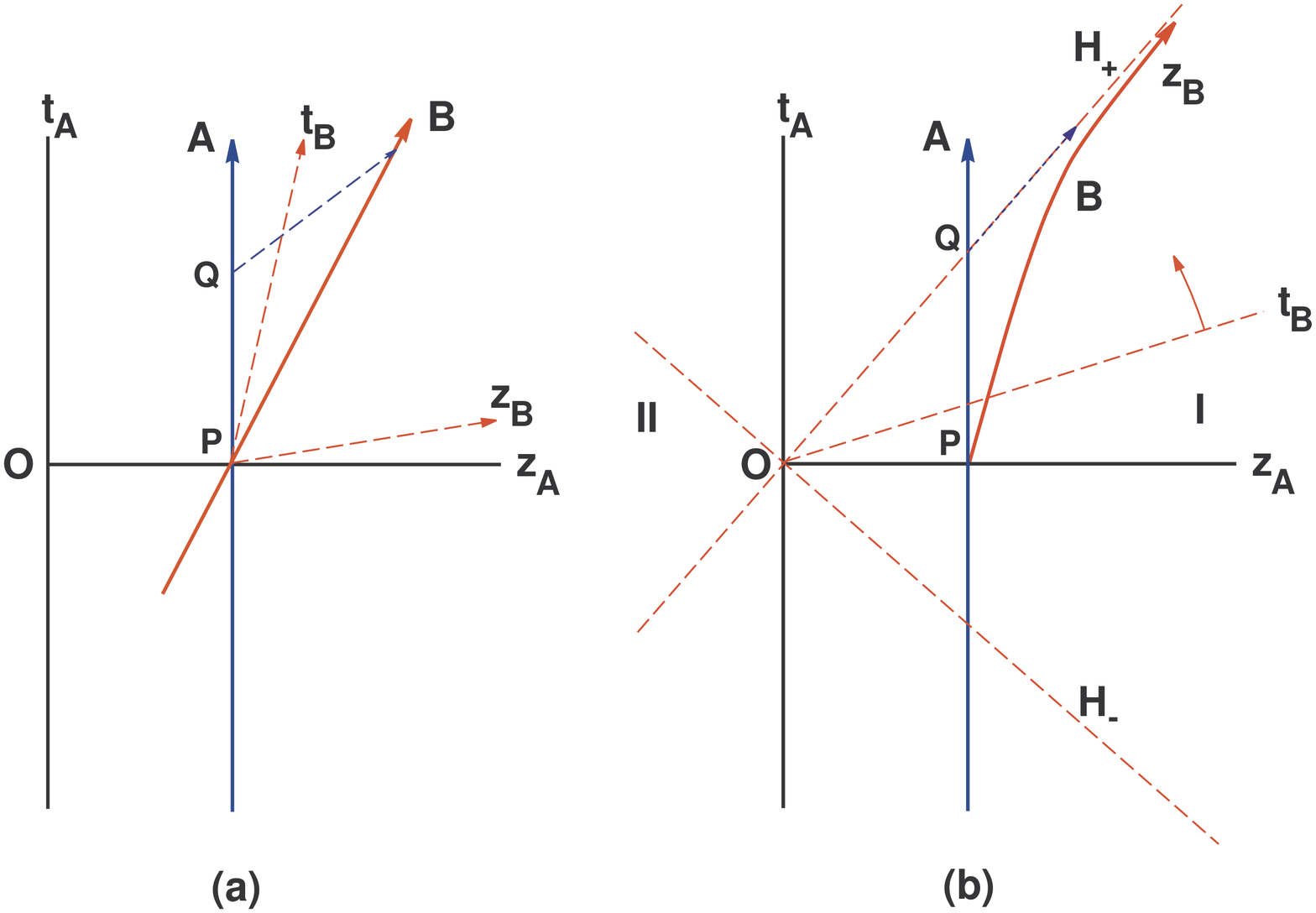,height=4.0in,width=6.0in}
\caption{(a) Minkowski diagram for the case of Alice (blue) stationary and Bob (red) travelling
at constant velocity. Alice and Bob share an entangled state $\Phi$ at the event $P$ (see text).
Alice can complete the teleportation protocol by sending classical signals to Bob
at a representative event $Q$. The entanglement fidelity of the state $\Phi$ is unaltered
if viewed from either Alice's or Bob's rest frame. (b) Alice (blue) is again stationary, but
Bob (red) undergoes constant acceleration. The light-like lines $H_-$ and $H_+$
form past and future particle horizon corresponding to Bob's proper times
$t_B = -\infty$ and $t_B = +\infty$ respectively. At the event $Q$ Alice crosses
$H_+$ (in her finite proper time $t_A$), and can no longer communicate with Bob.
Bob, however, can still send signals to Alice across $H_+$. The status of the entanglement
fidelity of the state $\Phi$ is unclear.}\label{accelframe}
\end{figure}

The case of entanglement for accelerated observers poses a whole host of new problems.
Consider first, for example the situation depicted in \Fig{accelframe}a in which Bob (red worldline)
is moving with momentum $\pv$ in the $z$ direction relative to a stationary Alice (blue worldline).
At the event $P$ let Alice and Bob share an entangled state $\Phi$, described by Alice as
\be{74}
\Phi = \frac{1}{\sqrt{2}} \,
\Big(
u_A(\mbf{0},1/2) \otimes u_B(\pv,1/2) + u_A(\mbf{0},-1/2) \otimes u_B(\pv,-1/2)
\Big).
\ee
If Alice has some other single particle state $\Psi$ which she wishes to teleport to Bob,
she can perform the usual procedure of mixing $\Psi$ with her portion of $\Phi$ and
transmit the result of her Bell measurement to Bob along a classical channel, depicted in
\Fig{accelframe}a as a light signal emitted at the event $Q$. If we consider the teleportation
from an inertial frame in which Bob is at rest, the situation is symmetric and Bob observes
the same state $\Phi$ except now the momentum for his particle is zero and for Alice it is $-\pv$.
Since we are boosting along the direction of motion of Bob, this is the trivial case of
zero Wigner rotation, so the spins are unaltered. Therefore the entanglement fidelity of $\Phi$
is unaffected, as we would expect.

The situation is very different if Bob is not travelling at constant velocity. Consider \Fig{accelframe}b
in which Bob (red worldline) is undergoing constant acceleration $a$, while Alice (blue worldline)
again remains stationary. Bob's coordinates $(z_B,t_B)$ are related to Alice's coordinates $(z_A,t_A)$ by
\be{75}
z_A = z_B \, \cosh a t_B, \qquad t_A = z_B \sinh a t_B.
\ee
In these \textit{Rindler} coordinates, Bob moves on a hyperbola of constant $z_B$, crossing
lines of his proper time $t_B$, which are straight (dotted red) lines emanating from the origin $O$.
At event $P$ Alice and Bob again share the entangled state $\Phi$, and Alice wishes to
teleport her state $\Psi$ to Bob.
Bob's world is very different from Alice's since he perceives that he is moving through
a thermal bath of radiation at the Unruh temperature $T_U=\hbar a / 2\pi k_B c$, where $k_B$ is
Boltzman's constant. Since Alice is in an inertial Lorentz frame, she perceives no such
Unruh radiation. In fact it is unclear how states would transform between Alice's and
Bob's reference frame since they each employ inequivalent quantization schemes \cite{birrel}.
Alice follows the usual quantization scheme in Minkowski spacetime, as discussed in
this paper, and her states are built up from the Minkowski vacuum $|0\rangle_M$ by
the usual Minkowski creation and annihilation operators $a^\dagger_M$ and $a_M$, such that
$a_M \, |0\rangle_M=0$. The right and left \textit{Rindler wedges} $z_A > 0, z_A > |t_A|$ and
$z_A < 0, |z_A| > |t_A|$ labelled $I$ and $II$ respectively in \Fig{accelframe}b, each support complete, and
distinct quantization schemes. This results in operators $\adag_I,\, a_I$ and $\adag_{II},\, a_{II}$
and vacua $|0\rangle_I$ and $|0\rangle_{II}$ in region $I$ and $II$ respectively,
inequivalent to each other and to $|0\rangle_M$.
The Rindler Hamiltonian $H_R$ which annihilates $|0\rangle_M$ and generates time translations
with respect to Bob's proper time $t_B$ is given by $H_R = H_I - H_{II}$, where for a fixed mode
$H_I \sim \adag_I\,a_I$ and $H_{II} \sim \adag_{II}\,a_{II}$. The Minkowski vacuum through
which Bob moves is described by a product over modes of maximally entangled two-mode squeezed states, comprised
of superpositions of Fock states of the form $|n\rangle_{I} \otimes |n\rangle_{II}$ for each mode.
(Note: a particle in the right Rindler wedge is correlated with an antiparticle in the left Rindler wedge
with opposite spatial momentum, and visa versa).
However, since Bob lives in region $I$,
he describes his physics in terms of states constructed solely from the operators $\adag_I,\,a_I$.
In addition, Bob is causally disconnect from region $II$, with the light-like lines $H_-$ and
$H_+$  in \Fig{accelframe}b acting as his past ($t_B=-\infty$) and future ($t_B=+\infty$) particles horizons.
Thus by tracing the maximally entangled state $|0\rangle_M\,\langle 0|$ over region $II$ states,
Bob describes the Minkowski vacuum by a maximally mixed, thermal reduced density matrix. A particle
detector carried by Bob will observe the unusual behavior of excitation of the detector accompanied by
the emission of a Minkowski particle, i.e. a particle registered by an inertial detector \cite{UnruhWald,Audretsch94}.

In examining \Fig{accelframe}b one sees that Alice's last communication with Bob is at the event $Q$ where
she crosses Bob's future particle horizon $H_+$. This occurs at $t_B=\infty$ with respect to Bob's
proper time, yet at some finite proper time with respect to Alice's inertial frame. Clearly at this stage the teleportation
protocol cannot continue. More importantly is the observation that Bob can still communicate with
Alice (say by photons) after she crosses $H_+$, but Alice can no longer communicate with Bob.
In this asymmetric situation, with states described by different quantization schemes, it is not
at all apparent if the entanglement fidelity of the shared state $\Phi$ is preserved.
These considerations are the subject of a future publication.

It is worthwhile to note that by the equivalence principle, the situation considered above
can essentially be considered as the local Lorentz description of a static observer around a black hole.
In the case of constant acceleration in Minkowski space the Unruh radiation ultimately stems from the force that
is keeping Bob in the state of constant acceleration. At fixed position outside a black hole, the static
observer must accelerate to stay in place and experiences a thermal flux of Hawking radiation analogous (though different)
to the Unruh radiation in Minkowski space. In both cases the presence of a horizon plays a central role
in the resulting radiation that is perceived. The question of entanglement across the horizon, and whether or
not unitary evolution still holds when a pure state falls behind the horizon and is apparently converted
into pure thermal radiation is still actively debated under the name of the "black hole information
loss" problem \cite{preskill}.

It is tantalizing to contemplate whether Unruh and/or Hawking radiation might be derived from
a quantum information theoretic point of view. As a heuristic consideration, note that the infinitesimal work $\Delta W$,
performed on a massive particle over its Compton wavelength $\lambda_c$ (the particle's characteristic length over which
we could consider it to be co-moving with a given inertial frame of constant velocity for a time
$\Delta t_A = \lambda_c/v(t_A)$) is given by $\Delta W = F dx = (m a)\,(h / m c) = (2\pi)^2 k_B T_U$. Up to
a numerical factor this is the energy associated with the thermal bath that Bob perceives as he
accelerates through the Minkowksi vacuum. This energy is the source of such processes discussed above whereby
a detector carried by Bob observes an excitation accompanied by an emission of a Minkowksi particle.
By Landauer's erasure principal \cite{plenio} there is an energy, and hence an entropy cost to erase information.
Might this energy absorbed by Bob be considered as going into the erasure of
the correlations in the pure state density matrix $|0\rangle_M\,\langle 0|$ for the Minkowski vacuum
through which he is accelerating, resulting in an entropy increase whose net effect is to create
the thermal vacuum which he perceives?
In addition, can the loss, in principle, of access to a quantum communication resource such as teleportation,
when Alice crosses Bob's future horizon $H_+$, be thought of in terms of erasure of information, and
an increase in entropy which is maximized by a thermal mixed state?
These considerations will be explored in a future publication.

\acknowledgements{The authors would like to thank G.J. Stephenson,
T. Goldmann,  T. Obsorne, M. Serna, and G. Brennen for many useful discussions and helpful comments.
P.M.A. would like to thank the Centre for Quantum Computer Technology at
the University of Queensland, where this work was performed.


\end{document}